\documentclass[sigconf,screen,nonacm,review=false,timestamp=false]{acmart}

\AtBeginDocument{%
  \providecommand\BibTeX{{%
    \normalfont B\kern-0.5em{\scshape i\kern-0.25em b}\kern-0.8em\TeX}}}

\usepackage{tabularray}
\usepackage{lipsum}
\usepackage{color}
\definecolor{Seashell}{rgb}{0.945,0.945,0.945}
\usepackage{mdframed}

\usepackage{colortbl}
\usepackage{hhline}
\usepackage{booktabs}
\usepackage{arydshln}

\usepackage{appendix}

\usepackage{orcidlink}

\begin{document}
\title{M2AR: A Web-based Modeling Environment for the Augmented Reality Workflow Modeling Language}
\thanks{Fabian Muff and Hans-Georg Fill | ACM 2024. This is the author's version of the work. It is posted here for your personal use. Not for redistribution. The definitive Version of Record was published in MODELS Companion '24 ACM/IEEE 27th International
Conference on Model Driven Engineering Languages and Systems September
22--27, 2024 Linz, Austria, https://doi.org/10.1145/3652620.3687779 | Financial support is gratefully acknowledged by the \href{https://www.smartlivinglab.ch/en/}{Smart Living Lab} funded by the University of Fribourg, EPFL, and HEIA-FR.}

\author{Fabian Muff \orcidlink{0000-0002-7283-6603}}

\affiliation{%
  \institution{University of Fribourg}
  \streetaddress{Bd. de Pérolles 90}
  \city{Fribourg}
  \country{Switzerland}
  \postcode{1700}
}
\email{fabian.muff@unifr.ch}
\author{Hans-Georg Fill \orcidlink{0000-0001-5076-5341}}
\affiliation{%
  \institution{University of Fribourg}
  \streetaddress{Bd. de Pérolles 90}
  \city{Fribourg}
  \country{Switzerland}
  \postcode{1700}
}
\email{hans-georg.fill@unifr.ch}

\renewcommand{\shortauthors}{Muff and Fill}

\begin{abstract}
This paper introduces M2AR, a new web-based, two- and three-dimensional modeling environment that enables the modeling and execution of augmented reality applications without requiring programming knowledge. The platform is based on a 3D JavaScript library and the mixed reality immersive web standard WebXR. 
For a first demonstration of its feasibility, the previously introduced Augmented Reality Workflow Modeling Language (ARWFML) has been successfully implemented using this environment. 
The usefulness of the new modeling environment is demonstrated by showing use cases of the ARWFML on M2AR.
\end{abstract}





\maketitle

\setlength{\fboxrule}{1pt} 
\setlength{\fboxsep}{0pt} 
    
\section{Introduction}
\label{sec:introduction}

Augmented reality (AR) has a significant impact on the merging of the physical and digital world~\cite{roo2017one}. It enhances the user's perception by superimposing visual information, including images, videos, or virtual three-dimensional (3D) objects, onto real-world settings in real-time~\cite{zhou2008trends,azuma1997survey}. Thus, AR uses computer vision techniques, as well as see-through displays or screens to display virtual information aligned with the real physical world~\cite{schmalstieg2016augmented}. 

AR relies on markers or detectors to identify the location and orientation of physical objects in the real world, allowing visual information to be accurately mapped onto them in a 3D environment. To make complex AR scenarios feasible in practical applications, it is also necessary to integrate external data sources, e.g., sensor data, as well as conditions, triggers, and state changes to process these data~\cite{muff2023domain}.

Despite the technological advances and price decrease of AR hardware in recent years~\cite{yin2021virtual}, the development of augmented reality applications continues to present significant difficulties. Unlike traditional desktop applications, augmented reality applications require consideration of a third dimension in addition to recognizing and mapping the real world. As a result, creating augmented reality applications requires in-depth knowledge and advanced programming skills. There are various platforms and APIs that can help in the development of AR applications, including, for example, Vuforia\footnote{\url{https://library.vuforia.com/}}, Apple ARKit\footnote{\url{https://developer.apple.com/augmented-reality/arkit/}}, Google ARCore\footnote{\url{https://developers.google.com/ar}}, or MRTK\footnote{\url{https://github.com/Microsoft/MixedRealityToolkit-Unity}}. Nevertheless, the development process remains highly complex. Several approaches in Model-Driven Engineering (MDE) and conceptual modeling have been proposed to reduce this complexity. These include textual modeling languages for the definition of 3D and AR applications such as those presented in \cite{ruminski2014dynamic, lenk2012model}, model-driven generation of 3D or AR modeling tools such as in \cite{ruiz_rube2020model, campos_l_opez2023model, wolter2015generating}, and visual modeling approaches for modeling or annotating process models with AR features such as in~\cite{grambow2021context,seiger2021holoflows}. Furthermore, commercial low-code/no-code tools such as UniteAR\footnote{\url{https://www.unitear.com/}} or Adobe Aero\footnote{\url{https://adobe.com/products/aero.html}} can facilitate the creation of very simple AR scenes. Last, 3D modeling tools for 3D object modeling, e.g., 3DARModeler~\cite{do20103darmodeler} or Designar~\cite{reipschl_ager2019designar}, allow for the modeling of complex 3D objects, or scenes. However, they are not well suited for the creation of complex scenarios which are not solely based on visual markers. 

To facilitate the no-code definition of complex AR scenarios, a domain-specific visual modeling language, implemented on a two-dimensional (2D) modeling tool, has been presented~\cite{muff2023domain}. However, 2D modeling tools are not well suited for modeling 3D workflows and environments. This is because it is very difficult to imagine coordinates and rotations without perceiving the 3D environment spatially.  

What is lacking is a modeling environment that is integrated with a 3D modeling platform for proper 3D object positioning. To fill this gap and facilitate the creation of AR applications, we propose a web-based, three-dimensional modeling environment. It allows both two- and three-dimensional modeling and generation of augmented reality applications without programming knowledge and its execution as an AR application. The platform is built on the meta$^2$-model introduced in~\cite{muff2021initial} and based on the 3D JavaScript library \textit{THREE.js}\footnote{\url{https://github.com/mrdoob/three.js/}} and the mixed reality immersive web standard \textit{WebXR}~\cite{jones2023webxr}. This enables the definition of various AR workflow scenarios in different domains with the 3D-enhanced domain-specific visual modeling language \textit{ARWFML}~\cite{muff2023domain}, such as assembly processes, maintenance tasks, or learning experiences.

The paper is structured as follows. Section~\ref{sec:related_work} discusses related approaches relevant for this paper. Section~\ref{sec:platform} introduces the new modeling environment for AR applications, including its design and architecture, as well as a demonstration that includes an illustrative use case. Section~\ref{sec:conclusion} concludes the paper. 

\section{Related Work}
\label{sec:related_work}

Several approaches have been explored in the past for the model-based generation of augmented reality applications. In a recent literature study, relevant papers at the intersection of conceptual modeling and model-driven engineering on the one side, and virtual- and augmented reality on the other side were analyzed \cite{muff2023past}. The result was the identification of different research streams, which will be briefly characterized in the following.

There exist four main streams that include different approaches. (1) Textual modeling languages for the definition of 3D and AR applications, e.g.,~\cite{ruminski2014dynamic,lenk2012model}. (2) Model-driven generation of 3D or AR modeling tools, cf.~\cite{ruiz_rube2020model,campos_l_opez2023model,wolter2015generating}. These are approaches to generate new AR or 3D modeling tools for specific modeling languages. (3) Visual process modeling tools for specific AR processes, e.g.,~\cite{grambow2021context,seiger2021holoflows}. These approaches are either for specific modeling languages such as BPMN, or for specific use cases, e.g., smart home processes. Furthermore, there are (4) no-code platforms or 3D modeling tools for defining simple AR objects and scenes, such as UniteAR, Adobe Aero, Apple's Reality Composer\footnote{\raggedright\url{https://developer.apple.com/documentation/realitykit/realitykit-reality-composer}}, 3DARModeler~\cite{do20103darmodeler}, or DesignAR~\cite{peffers2008design}. However, these tools are not suitable for creating complex AR scenarios. 

What has been missing is a visual 3D modeling approach that enables the definition and execution of AR applications without the need for programming skills. 
Therefore, we had introduced a new domain-specific modeling language called \textit{AR Workflow Modeling Language} (ARWFML) that allows visual modeling of augmented reality applications~\cite{muff2023domain}. The modeling language was demonstrated by a prototypical implementation of the language in a 2D modeling tool and by transforming the models into AR applications running in an AR execution engine. However, it was found that a 2D modeling tool was not suitable for defining AR applications, because AR is inherently based on 3D environments that are difficult to describe in 2D models.
Additionally, we found that extending one of the existing web-based modeling platforms, such as WebGME \cite{MarotiKKBVJLL14}, AToMPM \cite{SyrianiVMHME13}, or GLSP \cite{MetinB23a}, would not be feasible due to the significant changes required for user interaction in 3D environments. Furthermore, existing modeling platforms could not be directly used as an execution environment for AR applications running on modern devices but would also need to be considerably modified and extended.
As a result, this paper introduces a 3D-enhanced, web-based modeling environment for modeling and executing augmented reality applications.

\section{M2AR: Web-based Modeling Environment for AR Applications}\label{sec:platform}
The new 3D-enhanced, web-based modeling environment for modeling and executing AR applications will be denoted in the following as \textit{M2AR}. It extends traditional 2D modeling by adding a third dimension to models and is based on a modern technology stack for supporting recent AR devices. In this way, model elements can be more easily positioned in three dimensional space as required by the previously developed \textit{ARWFML} language~ \cite{muff2023domain}.

The artifact presented in this paper is the second iteration of the \textit{ARWFML} project following design-science research (DSR) methods, cf. \cite{peffers2008design}. In a first iteration, the modeling method was developed and implemented in the 2D metamodeling platform ADOxx~\cite{adoxx}, which however only supports 2D environments and is based on a technology stack that is not suitable for modern, web-based environments.

\subsection{Design and Development}\label{sec:design}

This section discusses the requirements, the conceptual design and the technical implementation of the \textit{M2AR} environment, as well as the \textit{ARWFML} implementation within this environment. 

\begin{figure}[h]
    \centering
\includegraphics[width=0.8\linewidth]{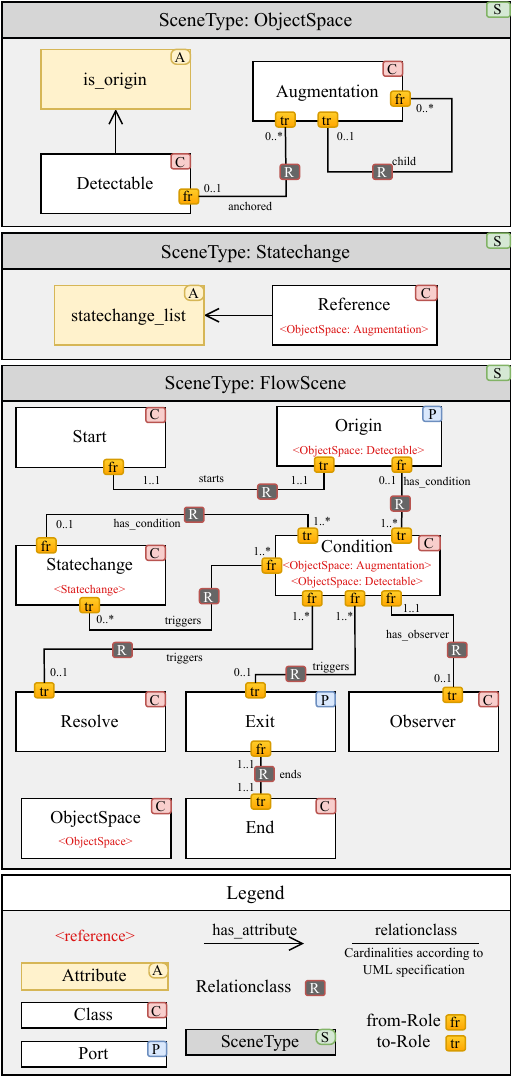}
    \caption{ARWFML metamodel for ObjectSpace, Statechange, and FlowScene.}
    \label{fig:mmar-metamodel}
\end{figure}

To derive the different requirements for the modeling environment, we first revisit the concepts of the \textit{ARWFML} language.
Figure \ref{fig:mmar-metamodel} shows the metamodel of three scene types \textit{ObjectSpace}, \textit{StateChange}, and \textit{FlowScene}. 
An \textit{ObjectSpace} is a model that defines both real-world and virtual objects required for the AR application. It consists of two main classes: \textit{Augmentation} and \textit{Detectable}. An \textit{Augmentation} is virtual data that can be used in AR, e.g., a 3D object or a text label. \textit{Augmentations} can be linked to other \textit{Augmentations} (\emph{child} relationclass) and can be linked to \textit{Detectables} through the \emph{anchored} relationclass. \textit{Relationclasses} are not connected directly to \textit{classes}, but to a \textit{from\_role} and a \textit{to\_role}. These \textit{roles} are then connected to other concepts, i.e., \textit{classes}. Thus, all the \textit{relationclasses} in Figure \ref{fig:mmar-metamodel} connect to two \textit{roles} denoted as \textit{fr} and \textit{tr}. A \textit{Detectable} is a representation of a real-world object, e.g., a 3D object or an image marker. \textit{Detectables} have an attribute ``is\_origin'' to indicate if they constitute a surrogate for the AR application's world origin. 
A \textit{Statechange} model describes changes in the appearance of \textit{Augmentations} defined in the \textit{ObjectSpace} model, such as visibility, position, or rotation. The \textit{Statechange} model includes only one class, \textit{Reference}, which lists the changes (statechange\_list) that occur in a referenced \textit{Augmentation}. This model is essential for representing dynamic changes in the AR environment. Note that \textcolor{red}{\textit{$<$reference$>$}} in the metamodel means that there is an attribute referencing to another instance of a Class, Port, or SceneType.
The \textit{FlowScene} model outlines the workflow of the AR application and its responses to various environmental conditions. It includes a \textit{Start} and an \textit{End} class, and contains a reference to an instance of the \textit{ObjectSpace} model (\textit{ObjectSpace} class).  The \textit{ObjectSpace} class has an \textit{Origin} port that refers to a \textit{Detectable} in the \textit{ObjectSpace} model. The \textit{FlowScene} also includes \textit{Statechanges} and \textit{Resolves}, which are triggered by \textit{Conditions}. \textit{Conditions} are triggered upon observing \textit{Augmentations} or \textit{Detectables}, or via the connected \textit{Observer} concept. \textit{Statechanges}, \textit{Conditions} and \textit{Resolves} are interconnected and determine how the resulting application shall respond to specific conditions in the AR environment, thereby guiding its workflow and interactions. Further details of the language are described in~\cite{muff2023domain}.

\begin{figure}[t]
    \centering    \includegraphics[width=.95\linewidth]{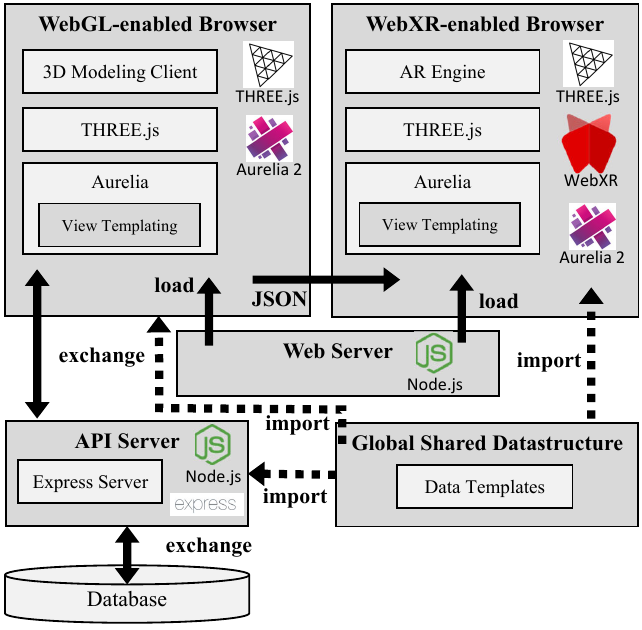}
    \caption{Architectural overview of the web-based modeling environment (\textit{M2AR}) for modeling AR applications.}
    \label{fig:architecture}
\end{figure}

Based on the \textit{ARWFML} concepts and the general objective of the solution, the following requirements (denoted as \textbf{R}) were derived.
\begin{itemize}
\item \textbf{R$_1$}: In order for the modeling platform to allow for the creation of \textit{ARWFML} models, it must support all defined \textit{ARWFML} concepts, as outlined in \cite{muff2023domain}. This includes the ability to model with the three different \textit{SceneTypes} on the same platform, interact with and manipulate 3D objects, and dynamically load external images and 3D objects. 
\item \textbf{R$_2$}: As \textit{ARWFML} is a 3D-based language, the modeling environment must allow for easy manipulation of the pose and orientation of 3D objects. 
\item \textbf{R$_3$}: To ensure platform independence and accessibility on various devices, such as smartphones, tablets, and head-mounted displays (HMDs), the environment should be web-based.  
\item \textbf{R$_4$}: The modeling environment shall be able to execute the created AR applications on different devices. Thus, to enable execution of models as an AR application, the platform must provide a module for executing the created models as an AR application.
\item \textbf{R$_5$}: The modeling environment must have the ability to store and access model data in a flexible manner. Thus, modules for exchanging and persisting model data are required.
\item \textbf{R$_6$}: The AR domain is characterized by a rapidly changing ecosystem, including advancements in tracking technology and evolving legal regulations. To ensure adaptability to new language requirements, the different parts of the modeling environment should be based on a common data structure that allows for flexible adaptation capabilities. 
\end{itemize}

Based on these requirements, a web-based modeling environment has been conceptualized and implemented. It is important to note that the introduced platform is independent of ADOxx used in the first iteration. Figure \ref{fig:architecture} shows the architectural overview of the \textit{M2AR} modeling environment with its different modules. This includes (1) a database server with \textit{PostgreSQL}\footnote{\url{https://www.postgresql.org/docs/}}, (2) an \textit{API Server} running as \textit{Node.js}\footnote{\url{https://github.com/nodejs/node}} application and \emph{express}\footnote{\url{https://github.com/expressjs/express}}, 
(3) a \textit{Node.js} web server providing (4) 
the web-page files for a \textit{3D Modeling Client} running \textit{Aurelia2}\footnote{\url{https://github.com/aurelia/aurelia}} and the \textit{JavaScript} \textit{WebGL} visualization framework \textit{THREE.js}\footnote{\url{https://github.com/mrdoob/three.js}}, and (5) the web-page files for an \textit{AR Engine} that takes JSON models from the \textit{3D Modeling Client} as input to execute modeled AR applications. The \textit{AR Engine} runs with the same technology stack in conjunction with the \textit{WebXR Device API}~\cite{jones2023webxr} enabling AR capabilities in the browser. These two web modules are separated, since the \textit{3D Modeling Client} should also run on non-AR devices. Furthermore, the \textit{AR Engine} could also run offline and independent of the \textit{3D Modeling Client} and the \textit{API Server}.
In addition, there is (6) a \textit{Global Shared Datastructure} module defining data templates, conforming to a previusly proposed meta$^2$-model~\cite{muff2021initial}, that can be used in all modules and allows for the flexible adaptation of the modeling language.

For the implementation of all components we used \textit{TypeScript}\footnote{\url{https://github.com/microsoft/TypeScript}}, which builds on \textit{JavaScript} by adding type checking during development and runs in all major browsers.

\subsection{Demonstration}\label{sec:demonstration}

For demonstrating the functionality of our prototype in terms of modeling and the execution of an AR application with the new \textit{M2AR} environment, we present in the following a language overview and a use case example. 



\subsubsection{Language Overview}

\begin{figure}[t!]
    \centering    \includegraphics[width=.9\linewidth]{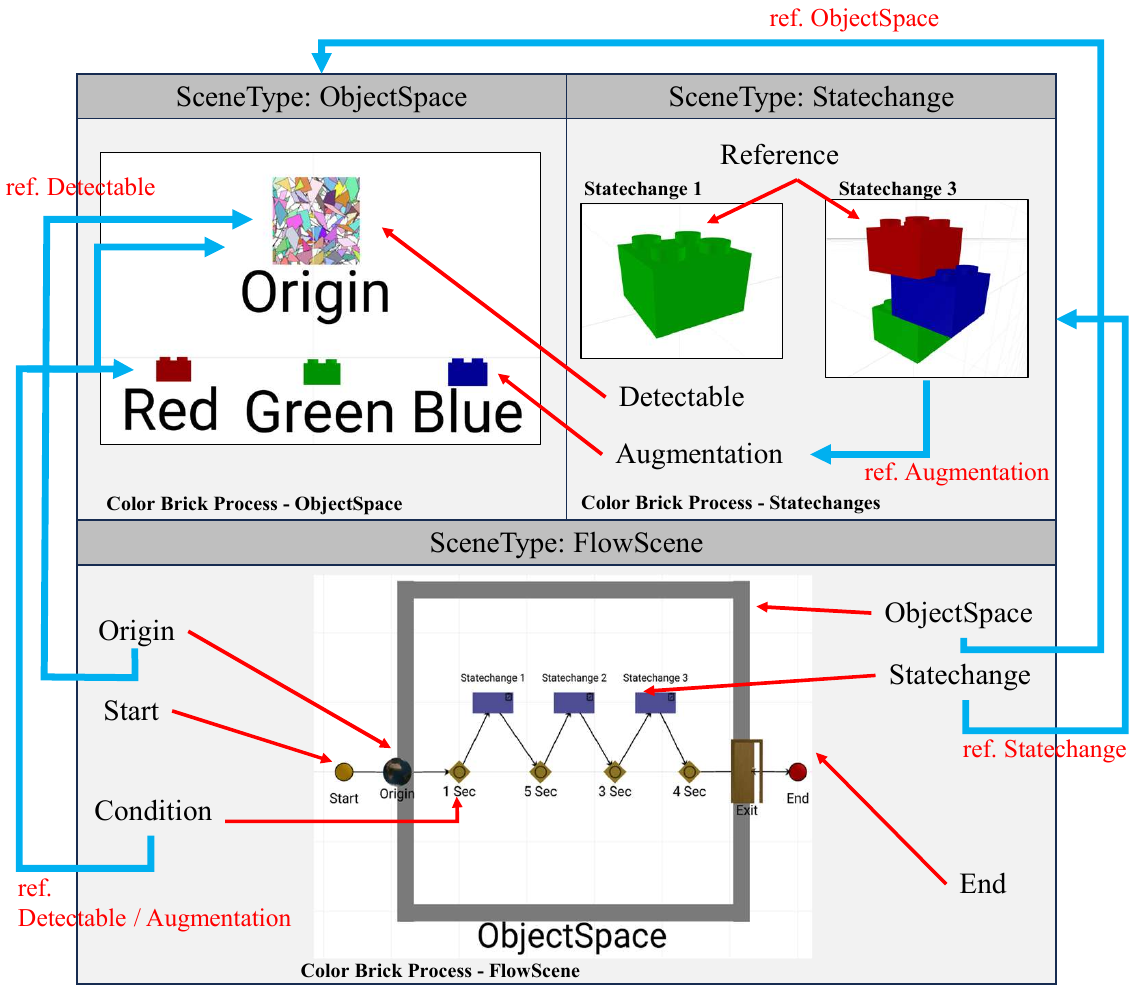}
    \caption{ARWFML language overview, demonstrated in an example of a color brick assembly process on the \textit{3D Modeling Client}.}
    \label{fig:language_overview_lego}
    \vspace{-.3cm}
\end{figure}

Figure \ref{fig:language_overview_lego} provides a language overview by showing a subset of the different models needed for a modeling use case. The \textit{ObjectSpace} model (top left) defines one \textit{Detectable} instance as image marker representation and three \textit{Augmentation} instances as 3D objects. The \textit{Statechange} area (top right) displays two examples of \textit{Statechange} model instances that define the appearance of the referenced \textit{Augmentations} when triggered. The \textit{FlowScene} (bottom) defines the actual path of the AR workflow. It references a \textit{Detectable} as the world-origin of the AR application, as well as different \textit{Conditions}, in this example four \textit{timer conditions}. It would also be possible for \textit{Conditions} to reference \textit{Detectables} in the form of markers or 3D objects, or \textit{Augmentations} as \textit{click condition}. Furthermore, the \textit{FlowScene} defines three \textit{Statechange} instances, referencing \textit{Statechange} models to be triggered.

The following sections provide a use case example for modeling with the \textit{M2AR Modeling Client} and executing the modeled AR applications in the \textit{AR Engine} to demonstrate the functionality of the ARWFML implementation on \textit{M2AR}.

\subsubsection{Color Brick Use Case}
The idea of this use case is to show a user in AR how to assemble colored bricks in the real world. Therefore, the resulting AR application should display the order and position of the over-scaled colored bricks in AR, and the user can then build the same structure with real colored bricks. In this use case, a green, a blue, and a red color brick are assembled.

The bottom of Figure \ref{fig:language_overview_lego} shows the \textit{FlowScene} model for the color brick use case. The \textit{FlowScene} model contains an \textit{ObjectSpace} class instance that references the previously created \textit{ObjectSpace} model. The \textit{ObjectSpace} class instance contains an \textit{Origin Port} that will reference a \textit{Detectable} from the \textit{ObjectSpace} model as the origin of the AR environment. Furthermore, the \textit{FlowScene} model defines \textit{Conditions} and \textit{Statechanges}. In this use case, there are four \textit{Conditions} as timer conditions and three \textit{Statechange} class instances that will reference to \textit{Statechange} models. 
The top left model in Figure \ref{fig:language_overview_lego} shows the \textit{ObjectSpace} model of the use case. The \textit{ObjectSpace} model contains one \textit{Detectable} that is marked as ``origin''. Furthermore, there are three \textit{Augmentations}. Each \textit{Augmentation} is visualized as green, blue, or red color brick. These representations are uploaded as GLTF\footnote{\url{https://www.khronos.org/gltf/}} file to the \textit{Object 3D} attribute of the \textit{Augmentations}.
The top right of Figure \ref{fig:language_overview_lego} shows two examples of \textit{Statechange} models in the 3D modeling view of the \textit{M2AR} \textit{3D Modeling Client}. As visible in the example, with the new 3D view it is possible to spatially align the referenced \textit{Augmentations}, and thereby define how the referenced \textit{Augmentations} should be visualized in AR during runtime when the \textit{Statechange} is triggered. In a last step, it must be ensured that all references are created, e.g., the \textit{Origin} reference as well as references for \textit{Conditions}, \textit{Statechanges}, and \textit{References}.   
When modeling is complete, the \textit{ObjectSpace}, the \textit{FlowScene}, and the three \textit{Statechange} models can be exported as JSON file through a button in the \textit{3D Modeling Client}.
    
\begin{figure*}[t!]
    \noindent 
    \begin{minipage}{0.32\textwidth}
    \fbox{\includegraphics[width=.99\linewidth]{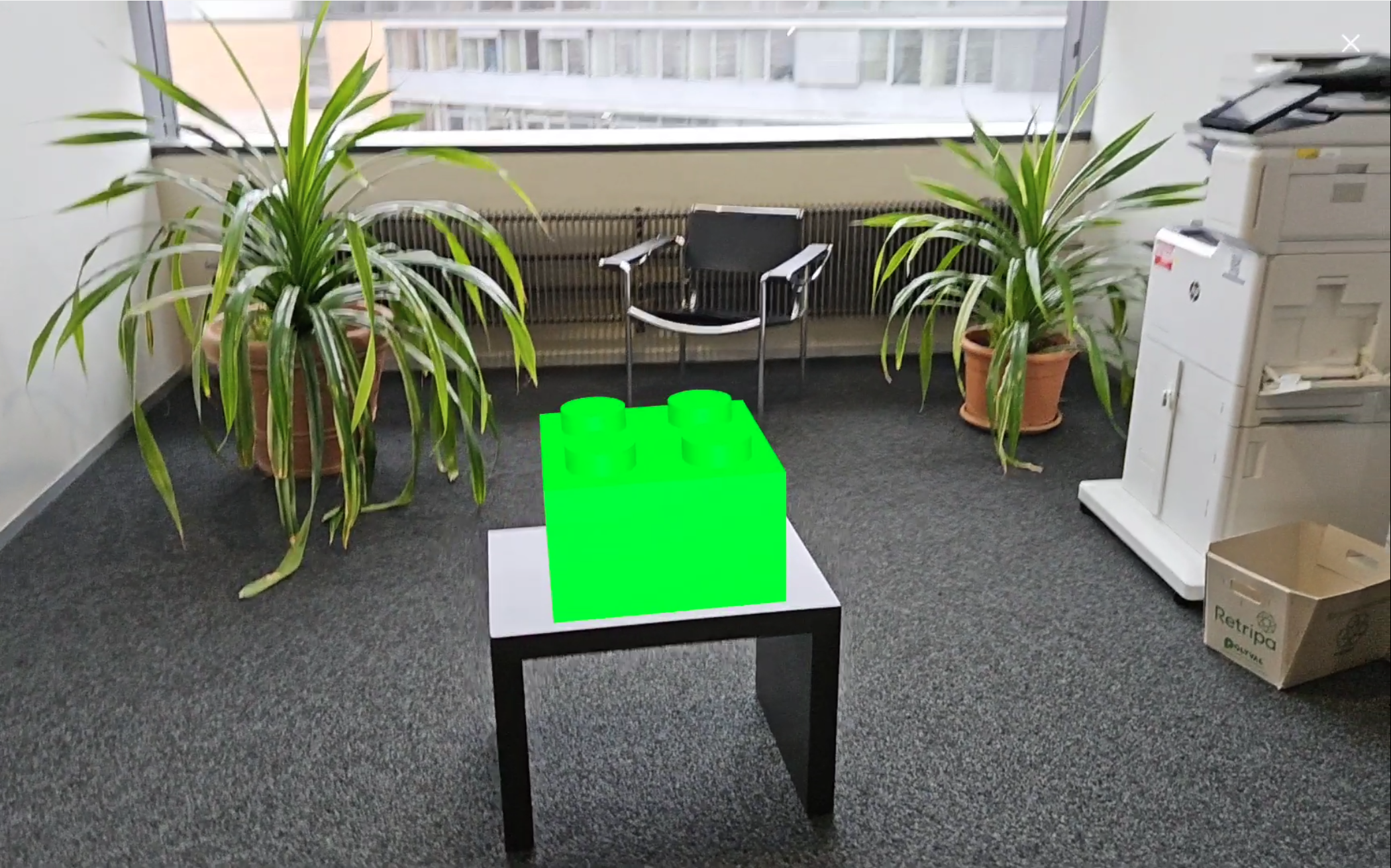}}
    \end{minipage}\hfill
    \begin{minipage}{0.32\textwidth}
      \fbox{\includegraphics[width=.99\linewidth]{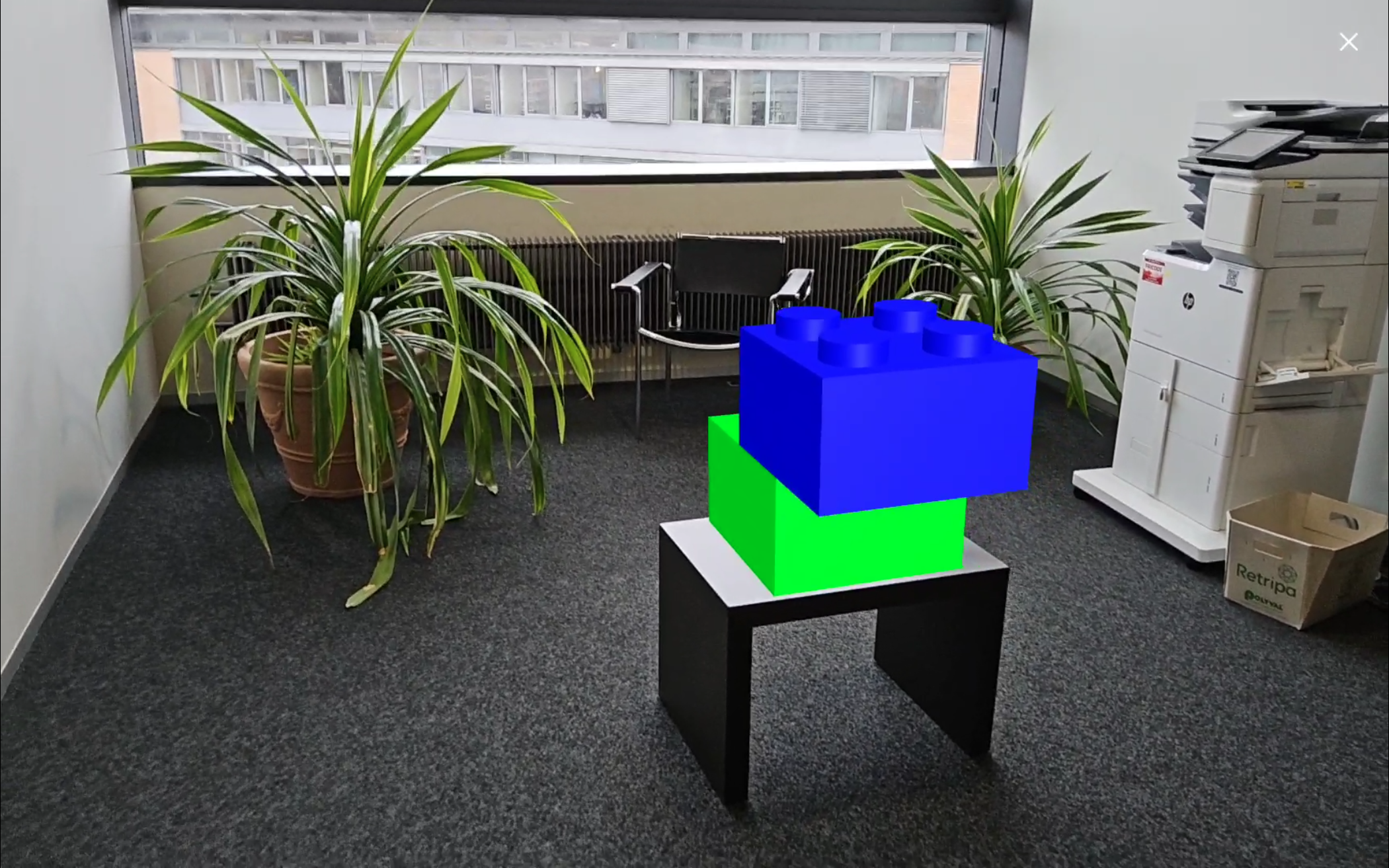}}
    \end{minipage}\hfill
    \begin{minipage}{0.32\textwidth}
      \fbox{\includegraphics[width=.99\linewidth]{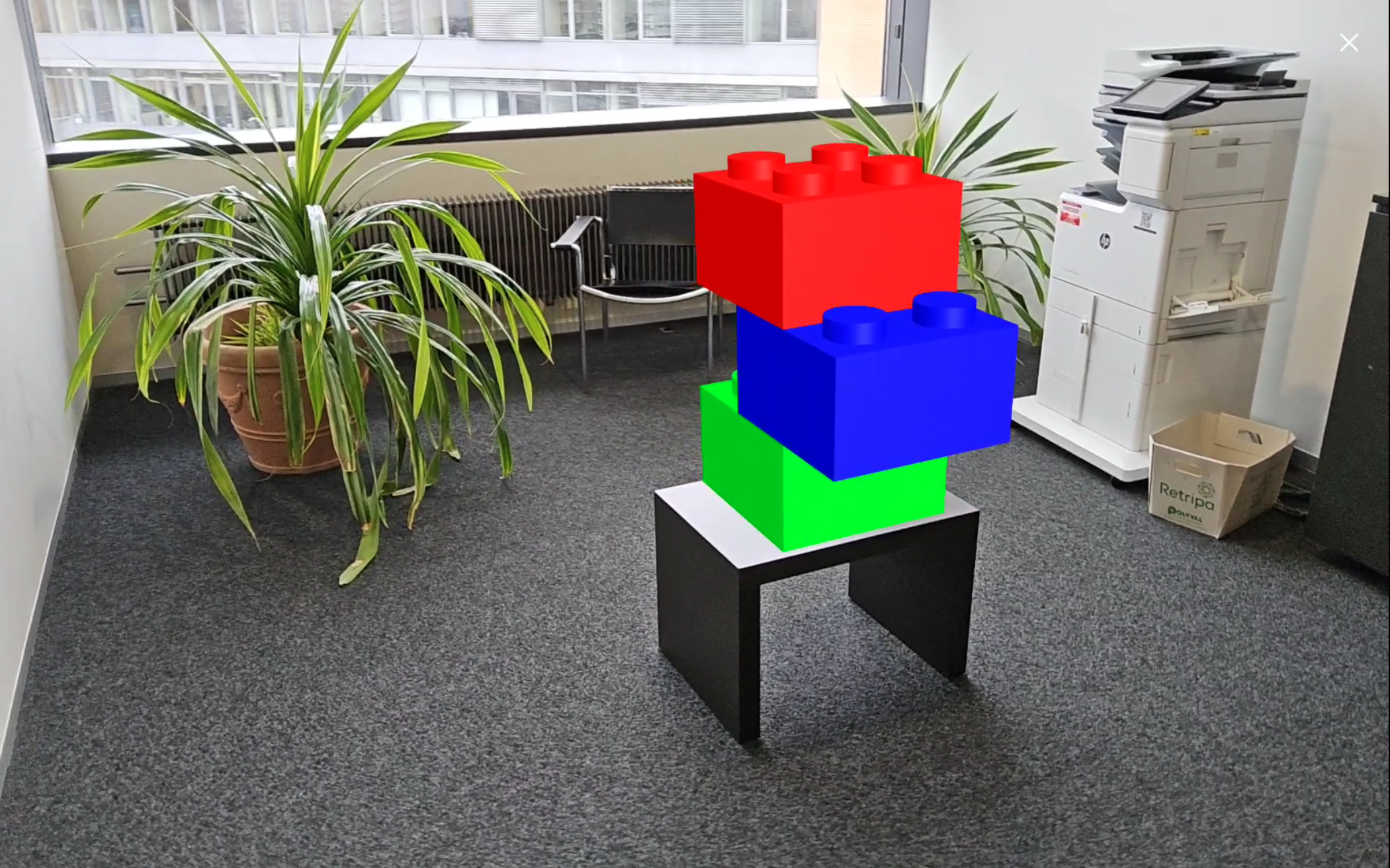}}
    \end{minipage}
    \caption{Screenshots of the color brick use case in the AR Engine, taken on a \textit{Samsung Galaxy Tab S7} tablet.}
    \label{fig:Lego_UseCase_Statechanges_engine}
\end{figure*}

The \textit{AR Engine} to run the exported models is accessible on any WebGL-enabled browser device supporting the \textit{WebXR device API}, e.g., smartphones, tablets, or AR HMDs. In the \textit{AR Engine}, one can upload the previously exported JSON file and start the AR application. 
As soon as the AR session is loaded, the \textit{AR Engine} checks for the detection of the origin. After the origin detection, the AR workflow starts according to the defined \textit{FlowScene}. Figure \ref{fig:Lego_UseCase_Statechanges_engine} shows three screenshots of all three \textit{Statechanges} in the running \textit{AR Engine} while executing the color brick use case defined above. The screenshots have been taken on a \textit{Samsung Galaxy Tab S7} tablet. Additional examples and videos are available online~\footnote{\url{https://doi.org/10.5281/zenodo.13268621}}.

\section{Conclusion and Outlook}
\label{sec:conclusion}
The definition of AR applications is a growing topic and is expected to become even more relevant with the global availability of the new \textit{Apple Vision Pro}\footnote{\url{https://www.apple.com/apple-vision-pro/}} in 2024. 

In this paper, we presented a new web-based, multi-dimensional, modeling environment (\textit{M2AR}) that allows the modeling and execution of augmented reality applications without programming knowledge. We achieved this by implementing the previously introduced \textit{ARWFML} \cite{muff2023domain} on a new web platform that is based on the 3D JavaScript library \textit{THREE.js} and the mixed reality immersive web standard \textit{WebXR}~\cite{jones2023webxr}, allowing the flexible definition of different AR workflow scenarios. These workflows can then be executed with an introduced \textit{AR Engine} on any WebXR supported device. 

The prototypical implementation of the \textit{M2AR} environment was demonstrated and a use case of creating \textit{ARWFML} models on \textit{M2AR} was shown. In a next step, we will evaluate the \textit{M2AR} environment against the requirements derived at the beginning of the DSR cycle as well as with an empirical user study to assess the comprehensibility of the language and tool usability.

\bibliographystyle{ACM-Reference-Format}
\bibliography{bibliography}

\end{document}